\documentclass{PoS}

\usepackage{amsmath}
\usepackage{lineno}

\title{Exploiting the time of arrival of Cherenkov photons at the 28 m H.E.S.S. telescope for background rejection: Methods and performance.}

\ShortTitle{Exploiting the timing information in H.E.S.S.}

\author{\speaker{Rapha\"el Chalm\'e-Calvet}\\ 
        LPNHE - Universit\'e Pierre et Marie Curie\\
        E-mail: \email{raphael.chalme-calvet@lpnhe.in2p3.fr}}

\author{Markus Holler\\
        LLR - Ecole Polytechnique\\
        E-mail: \email{markus.holler@llr.in2p3.fr}}

\author{Mathieu de Naurois\\
		LLR - Ecole Polytechnique\\
		E-mail: \email{denauroi@in2p3.fr}}
	
\author{Jean-Paul Tavernet\\
		LPNHE - Universit\'e Pierre et Marie Curie\\
		E-mail: \email{tavernet@lpnhe.in2p3.fr}}

\abstract{	
	In 2012, the High Energy Stereoscopic System (H.E.S.S.) was expanded by a fifth telescope (CT5). 
	With an effective mirror diameter of $28\,$m, CT5 is able to detect the Cherenkov light of very faint gamma-ray air showers, thereby significantly lowering the energy threshold of this telescope compared to the other four telescopes. 
	Extracting as much information as possible from the recorded shower image is crucial for background rejection and to reach an energy threshold of a few tens of GeV.
	The camera of CT5 is conceived to register the time of the charge pulse maximum with respect to the beginning of the 16 ns integration window of each pixel. 
	This information can be utilised to improve the event reconstruction. It also helps to reduce the background contamination at low energies. 
	We present new techniques for background rejection based on CT5 timing information and evaluate their performance.}

\FullConference{The 34th International Cosmic Ray Conference,\\
		30 July- 6 August, 2015\\
		The Hague, The Netherlands}

\makeatletter
\newcommand{\thickhline}{%
	\noalign {\ifnum 0=`}\fi \hrule height 2pt
	\futurelet \reserved@a \@xhline
}
\makeatother

\begin{document}
	

\section{Introduction}

Within the last few decades, Imaging Atmospheric Cherenkov Telescopes (IACTs) have greatly enhanced the knowledge about the most violent processes in the universe. Using the Earth's atmosphere as a calorimeter leads to a large detection area, which enables IACTs to study sources at $\gamma$-ray energies from several tens of GeV up to around $100\,$TeV. The disadvantage of this approach is the dominant background which mainly corresponds to charged cosmic rays. Developing efficient mechanisms to suppress this component is thus crucial to improve the sensitivity of IACTs.
	
Background suppression in ground-based gamma-ray astronomy is generally based on the characterisation of the shower image that is recorded by the camera of an IACT. However, several efforts were carried out to make use of the arrival time information of the Cherenkov photons as an additional, independent variable. In $2007$, \cite{2007_Mazin} exploited the timing information of the first MAGIC telescope using a template-based reconstruction method that resembles the one of \cite{model}. Two years later the MAGIC collaboration presented several ways to improve the single-telescope performance by introducing several time-based variables \cite{2009_MAGIC_Timing}.

	H.E.S.S. is a system of five IACTs located in Namibia. 
	From $2004$ to $2012$ (corresponding to H.E.S.S. phase I), the system consisted of four identical IACTs (named CT1-4) with an effective mirror diameter of $12\,$m each, disposed on the corners of a square with $120\,$m edge length. 
	In 2012, the construction of an additional large-size (32 $\times$ $24\,\mathrm{m}$) telescope (CT5) in the centre of the array was completed, marking the beginning of H.E.S.S. phase II.

	The improved electronics of the CT5 camera can trigger events with a rate of up to $5\,$kHz. 
	This high trigger rate combined with its greater resolution 
	allows to have a performing monoscopic analysis using only the CT5 images. 
	In addition to the intensity, the $2048$ camera pixels record temporal informations of the events, such as the time over a given threshold and the Time of Maximum. 
	
	In this contribution, we present the status of the background rejection study using the Time of Maximum information of the CT5 camera. 
	The variables commonly used by the Model reconstruction \cite{model} are firstly reminded. 
	Then the Time of Maximum, its calibration and its potential for background suppression are described. 
	Finally, preliminary results of the approach as tested on Monte Carlo simulations and data of the AGN PKS 2155$-$304 are shown. 
		
\section{Usual variables for background rejection}

	In this section the principal methods of background rejection used within the H.E.S.S. monoscopic Model reconstruction are summarized. 

\subsection{Shape cuts}

	After the shower reconstruction, some primary cuts, called shape cuts, are performed. 
	They are not especially background rejection cuts but more similar to safety cuts allowing to reject non-physical or hardly analysable reconstructed shower images. 
	They also contribute to considerably reduce the number of events to analyse. 
	Here are presented the main shape cuts used by the Model analysis : 

\vspace{-2mm}

\begin{itemize}
	
	\item \textit{Minimum charge} : the minimum number of photo-electrons accepted in the image.
	
	\vspace{-2mm}
	
	\item \textit{Nominal distance} : distance between the centre of gravity of the shower and the camera center. This allow to only keep events for which the center of gravity is inside the camera.
	
\end{itemize}

\subsection{Goodness of fit}\label{Goodness_section}

	We describe here the shower goodness which is the main hadron rejecter of the Model analysis and attests of the quality of the model adjustment.
	For more details on its derivation, see \cite{model}. 
	The goodness $G$ of a $N$ pixel image is given by :

	\begin{equation} \label{Goodness}
		G = \frac{\sum\limits_{\text{i = 0}}^N[\ln L(s_i|\mu_i)-<\ln L>|_{\mu_i}]}{\sqrt{N}}
	\end{equation}
	where $s_i$ and $\mu_i$ are respectively the recorded signal in photoelectrons and the model derived from the reconstruction fit for a given pixel $i$.
	
	The likelihood $L(s|\mu)$ defined by the probability $P(s|\mu,\sigma_p,\sigma_\gamma)$ of having a signal $s$ when knowing a model $\mu$ is given by :
	\begin{equation}\label{Likelihood}
		\begin{split}
			\ln L(s|\mu) &= -2 \ln P(s|\mu,\sigma_p,\sigma_\gamma)\\	
			&= -2 \ln \left[\sum\limits_n \frac{\mu^n e^{-\mu}}{n!\sqrt{2\pi (\sigma_p^2 + n \sigma_\gamma^2)}}\exp\left(-\frac{(s-n)^2}{2(\sigma_p^2 + n \sigma_\gamma^2)}\right) \right]
		\end{split}
	\end{equation}
	where $n$ is the number of photoelectrons, $\sigma_p$ is the width the pedestal and $\sigma_\gamma$ is the width a the single photoelectron peak.
	
	Finally the average of the likelihood is : 
	
	\begin{equation}\label{Mean Likelihood}
		<\ln L>|_{\mu} = \int \ln L(s|\mu) \times P(s|\mu,\sigma_p,\sigma_\gamma) \,  \mathrm{d}s
	\end{equation}
	
\subsection{Primary depth}

	The primary depth is the atmospheric depth of the first Cherenkov emission of a shower. 
	It is reconstructed during the model analysis and allows to discriminate gamma from hadrons but also electron showers. 
	Indeed, the Cherenkov emission of electron showers is expected to begin higher in the atmosphere.

\begin{figure}[tbhp]
	\center
	\includegraphics[width=0.37\textwidth]{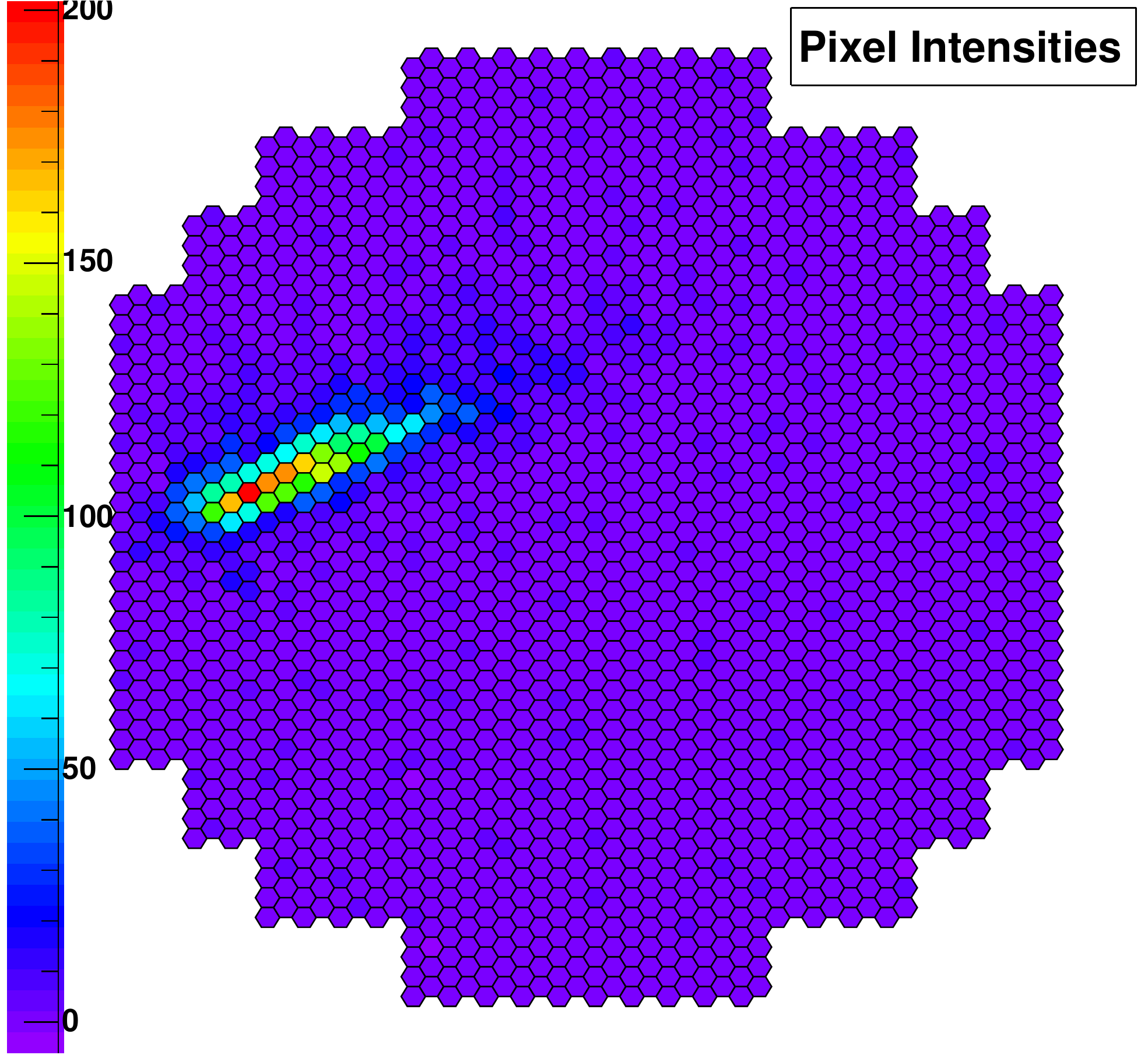} \hspace{15mm}
	\includegraphics[width=0.37\textwidth]{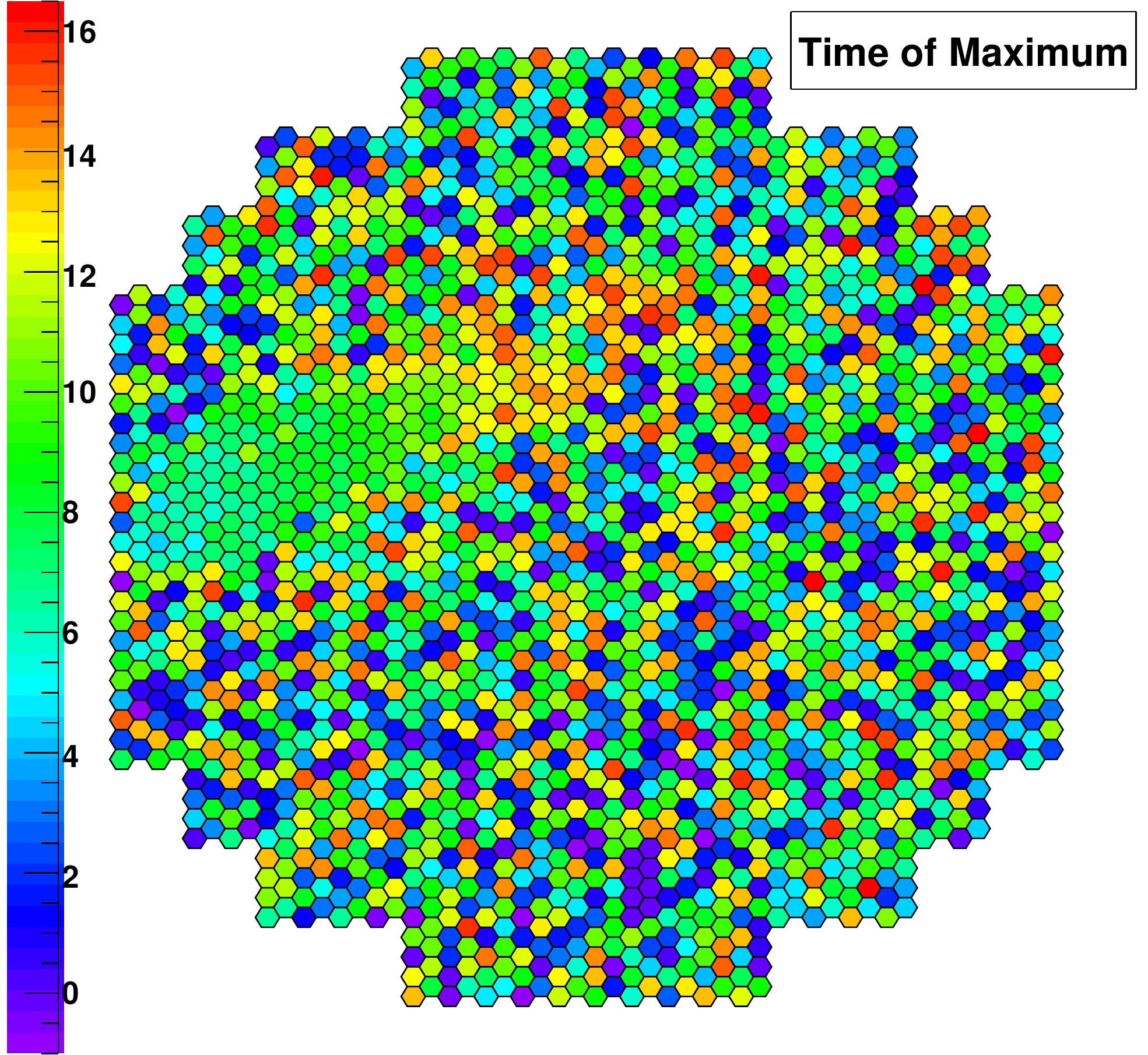}
	\caption{
		\label{ShowerEventDisplay}
		Shower candidate in the CT5 camera.
		The intensity display, on the left, shows the shower position in the camera (high intensity pixels) and the distribution of pixels containing only NSB photons (intensity close to 0 photon-electron).
		On the right, the Time of Maximum display presents a clear difference between the NSB photons with a ToM randomly distributed and the Cherenkov photons from the shower presenting a ToM gradient within the camera.
	} 
\end{figure}

\subsection{Night Sky Background Goodness}\label{NSBG_section}

	The Night Sky Background (NSB) Goodness is the most powerful cut in the case of monoscopic analysis. 
	Its main goal is to determine if the recorded image is compatible with an image which doesn't contain Cherenkov photons, but mostly NSB photons which will enlarge the pixel distribution of the base line position of the electronic pedestal.

	The NSB Goodness ($G_{NSB}$) is very comparable in its construction to the goodness of fit, presented in the section \ref{Goodness_section}. 
	The main difference is that we considerate a null model $\mu$:

	\begin{equation} \label{eq_NSBG}
		G_{NSB} = \frac{\sum\limits_{pix}\left(\ln L(s|\mu = 0)-<\ln L>|_{\mu = 0}\right)}{\sqrt{N}}
	\end{equation}
	where the likelihood definitions are identical to the equations (\ref{Likelihood}) and (\ref{Mean Likelihood}) by replacing $\mu$ by 0.

	In the monoscopic analysis, this variable rejects around 90\% of the events allowing to produce well normalized map and high signal over background ratio and significance. 
	Most of the rejected showers are low energy events, due to their low intensity which is more compatible with null model. 
	Thus, cutting harder on the NSB goodness, directly increases the energy threshold of the analysis.

	In order to keep the low energy events without degrading the analysis, other alternatives should be considered. 
	We present in the next sections one of them, using the timing information recorded by CT5.

\section{The temporal information in H.E.S.S.}

	The 2048 pixels of the fifth H.E.S.S. camera record two temporal informations within the 16 nanoseconds of the integration window. 
	The first one is the Time over Threshold which represents the duration of the electronic pulse over a given charge threshold. 
	The second one is the Time of Maximum which will be described into details in the following sections.

\subsection{Time of Maximum}

	During the integration window, the temporal information of the pulse is sampled every nanoseconds and each pixel records the temporal position of the pulse maximum, called Time of Maximum (ToM). 
	For an atmospheric shower, the ToM is expected to be very close from pixel to pixel, when for the pixels containing only NSB, it will be randomly distributed.
    The difference between the Cherenkov and NSB photons ToM distribution is well illustrated in the figure \ref{ShowerEventDisplay} showing an example of a shower like event in the CT5 camera.

\subsection{Calibration of the Time of Maximum}

	A calibration of the ToM over the camera is necessary because the 2048 pixels are distributed over 256 redout cards, which are not perfectly temporally synchronised.
	It is effectuated using Flat Field runs.
	
	During these runs, a laser flash illuminates uniformly the entire CT5 camera. 
	Due to the parabolic mount of the CT5 mirrors, the laser photons are expected to arrive in all pixels in the same time. 
	This duration is well smaller than the electronics resolution. 
	Comparing the Time of Maximum of the different pixels allows to reduce the disparities which can occur between them. 
	
	After calibration the Root Mean Square (RMS) of the ToM distribution of all the pixels goes from 0.6 ns to 0.02 ns.

\begin{figure}[htbp]
	\center
	\includegraphics[width=0.41\textwidth]{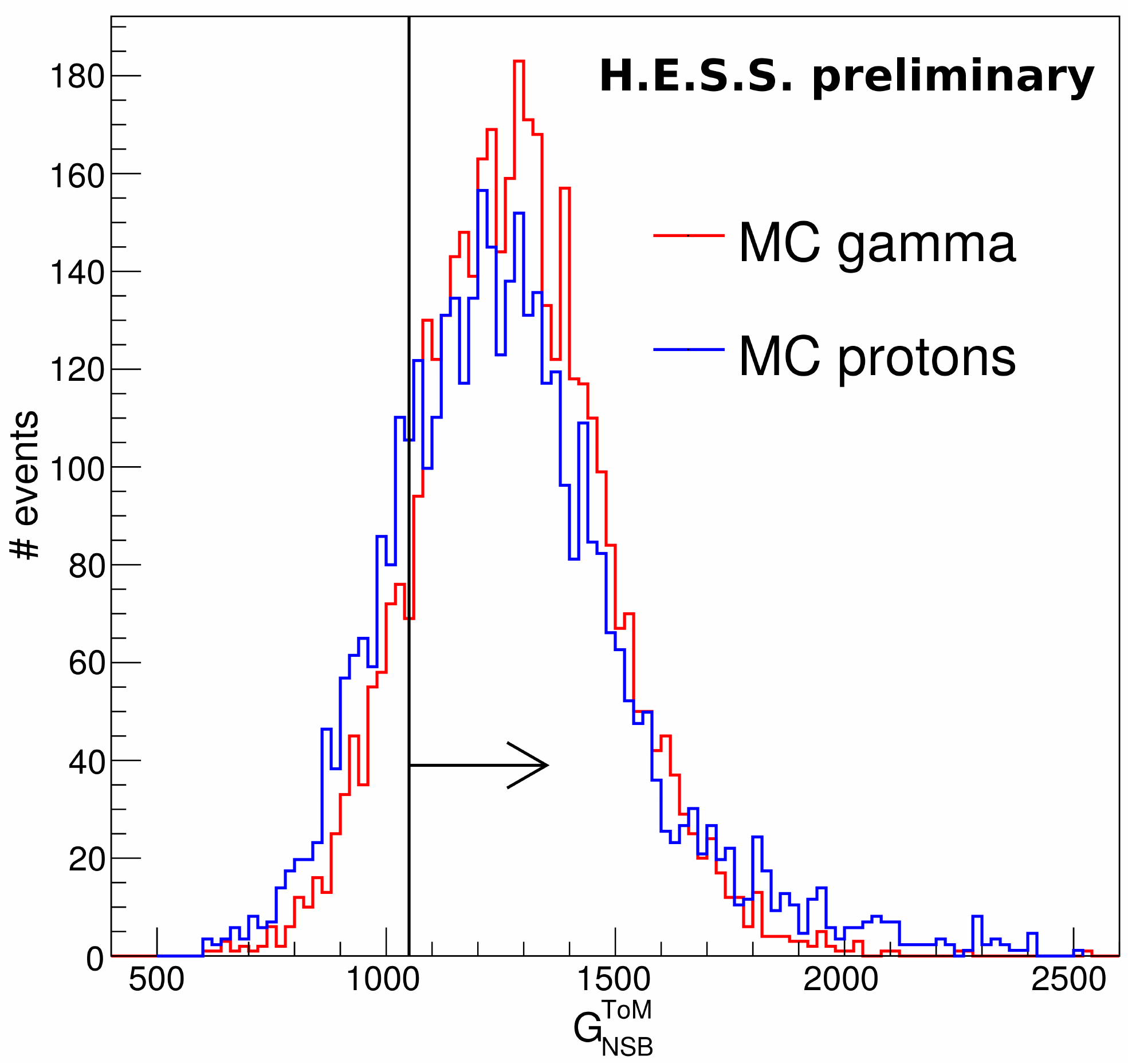}
	\includegraphics[width=0.58\textwidth]{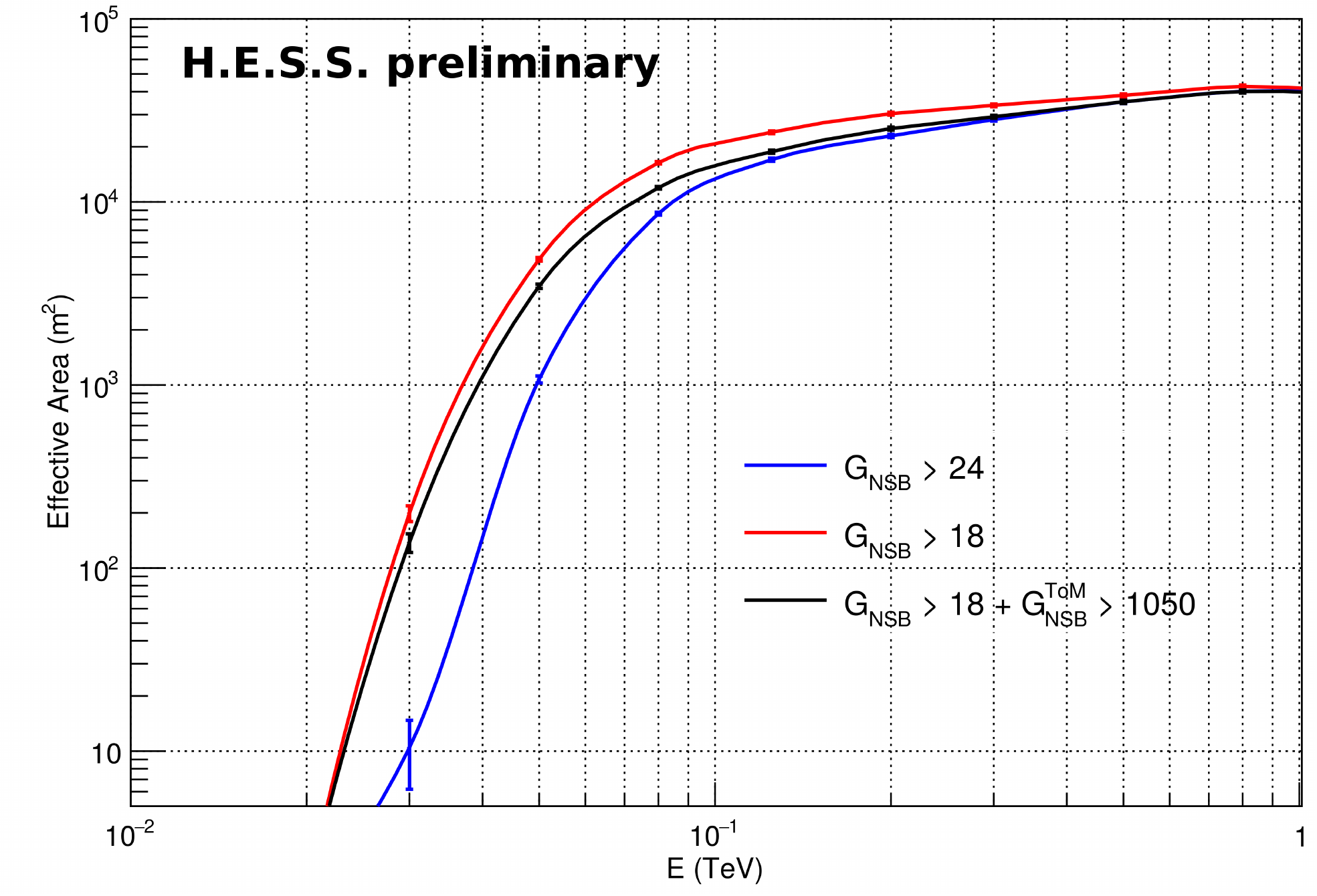}
	\caption{
		\label{MC distributions}
		Left: Distributions of $G_{NSB}^{ToM}$  for simulated protons (in blue) and gamma photons (in red) passing official H.E.S.S. cuts and $G_{NSB} > 18$.
		Right: Effective area for the three studied cut configurations at 18$^{\circ}$ zenith angle and source simulated at 1$^{\circ}$ off-axis from the camera center.
	}
\end{figure}

\section{Timing variables for background rejection}

	The loosest official cut configuration applied in the Model analysis involves a NSB Goodness cut ($G_{NSB}$, see section \ref{NSBG_section}) at a minimum of 24.
	Cutting at lower $G_{NSB}$, allows to gain a non negligible amount of low energy events but also introduce large systematics effect on the background rejection, especially in the map productions.
	The timing variable presented in this section aims to lower the energy threshold of the H.E.S.S. experiment while keeping a controlled background subtraction.

\subsection{Temporal Night Sky Background goodness}

	The conception of the temporal NSB goodness is very similar to its homonym in intensity (section \ref{NSBG_section}).
	Therefore, the definition of the ToM NSB goodness ($G_{NSB}^{ToM}$) is almost identical to the equation (\ref{eq_NSBG}) :
	
	\begin{equation} 
		G_{NSB}^{ToM} = \frac{\sum\limits_{i=0}^N\left(\ln L(ToM|\tau)-<\ln L>|_{\tau}\right)}{\sqrt{N}}
	\end{equation}

	Here we assume that the probability that a pixel has a Time of Maximum value $ToM$ when a model $\tau$ is expected is given by a Gaussian centered in $\tau$ with a width $\sigma$ equal to the temporal resolution of a pixel obtained during the calibration. 
	The likelihood is then given by :
	\begin{equation} 
		\ln L(ToM|\tau) = \ln(2\pi) + \ln(\sigma^2) + \frac{(ToM-\tau)^2}{\sigma^2}
	\end{equation}
	and the average likelihood by :
	\begin{equation} 
		<\ln L>|_{\tau} = \ln(2\pi) + \ln(\sigma^2) + 1
	\end{equation}
	
	The goal is to compare the recorded image to an image that doesn't contain Cherenkov photon. 
	From a temporal point of view, the ToM of a null image is randomly distributed in the camera.
	The model $\tau$ is thus picked randomly in a uniform distribution between 0 and 15 ns, in which is added a Gaussian temporal resolution.

	The loosest cut used in H.E.S.S. is obtained with the intensity NSB Goodness $G_{NSB}>24$. In the next sections, we will lower it to $G_{NSB}>18$ and study the behaviour of the temporal NSB Goodness. 
	This $G_{NSB}$ value has been taken to decrease sufficiently the energy threshold without increasing too much the map systematic uncertainties.

\subsection{Study of Monte Carlo simulation\label{MC section}}

	From the definition of the $G_{NSB}^{ToM}$, its main purpose is to reject NSB triggered events and gamma candidate shower which are too much contaminated with NSB photons.
	In addition, this section will show that this variable also allows to improve the hadron suppression of the Model monoscopic reconstruction.	
	The left of the figure \ref{MC distributions} presents the $G_{NSB}^{ToM}$ distributions for simulated gamma and proton events passing the usual H.E.S.S. cut and $G_{NSB}> 18$.
	Protons present a higher rate than gamma at low $G_{NSB}^{ToM}$, meaning that the proton shower timing dispersion is higher than gamma shower, as expected.
	Thus, a cut at  $G_{NSB}^{ToM}>1050$, limit where the gamma rate becomes greater than the proton rate, should improve the background rejection, by suppressing more protons than gamma.
	
	The right of the figure \ref{MC distributions} shows the effective area graphics as function of the simulated energy for the three studied configuration : 
	the loosest H.E.S.S cut ($G_{NSB}>24$), a drop in NSB Goodness ($G_{NSB}>18$) and this last drop adding the temporal NSB Goodness cut found in the previous paragraph ($G_{NSB}>18 + G_{NSB}^{ToM}>1050$).
	This plot shows that a rise of the NSB Goodness directly increases the energy threshold, while applying the $G_{NSB}^{ToM}$ cut has much less impact on the energy threshold.

\begin{table}[tbhp]
	\centering
	\begin{tabular}{|c|c|c|c|c|}
		\hline
		& N$_\sigma$ & S/B & N$_\gamma$ &	 Width of the OFF \\
		&  &  &  & significance distribution \\
		\thickhline
		$G_{NSB} >$ 24  & 18.4 & 0.39 & 1070 & 1.0 \\
		\hline
		$G_{NSB} >$ 18  & 17.2 & 0.21 & 1634 & 1.3 \\
		\thickhline
		$G_{NSB} >$ 18 + $G_{NSB}^{ToM} >$ 1050  & 13.6 & 0.24 & 899 & 1.1 \\
		\hline
	\end{tabular}
	\caption{\label{PKS table}Results of the PKS 2155-304 analysis for the three studied cut configurations. N$_\sigma$ is the significance of the source, N$_\gamma$ is the number of excess events and S/B is the signal over background ratio.}  
\end{table}

\begin{figure}[tbhp]
	\center
	\hspace{2mm}\includegraphics[width=0.325\textwidth]{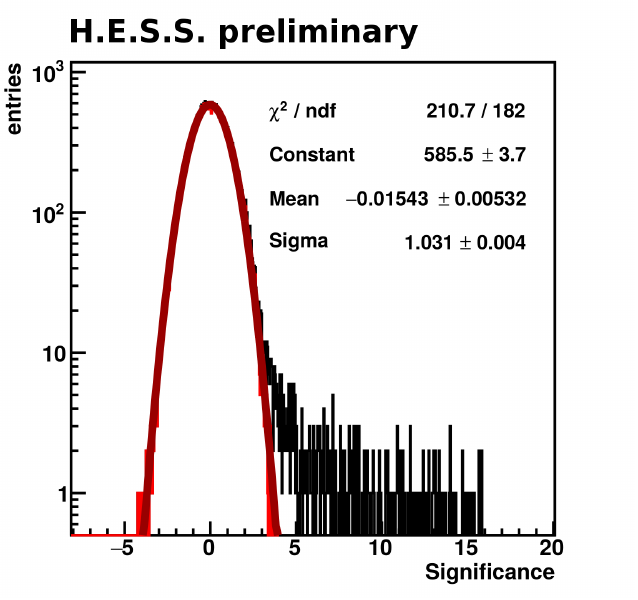}
	\includegraphics[width=0.325\textwidth]{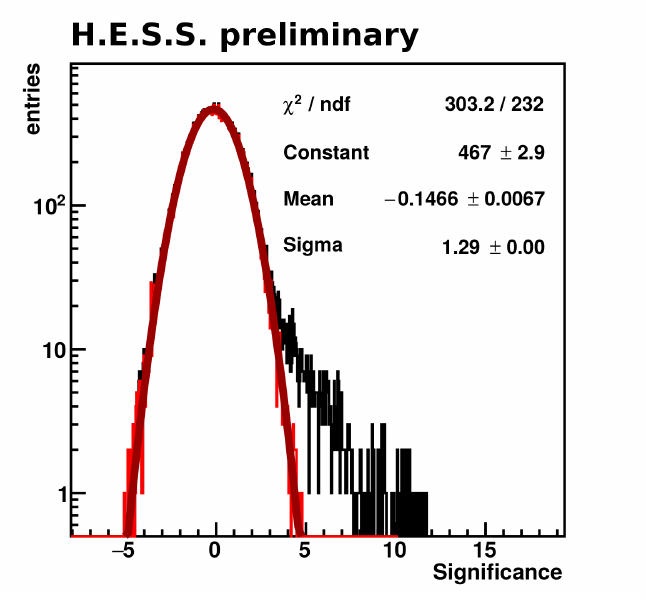}
	\includegraphics[width=0.325\textwidth]{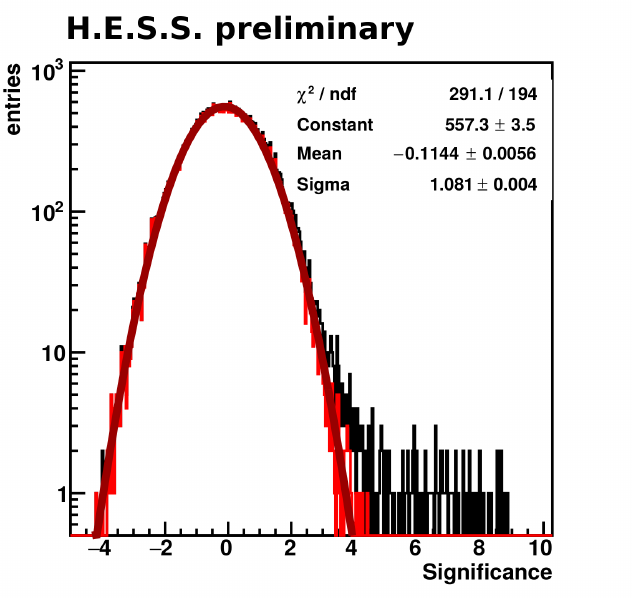}
	
	\includegraphics[width=0.325\textwidth]{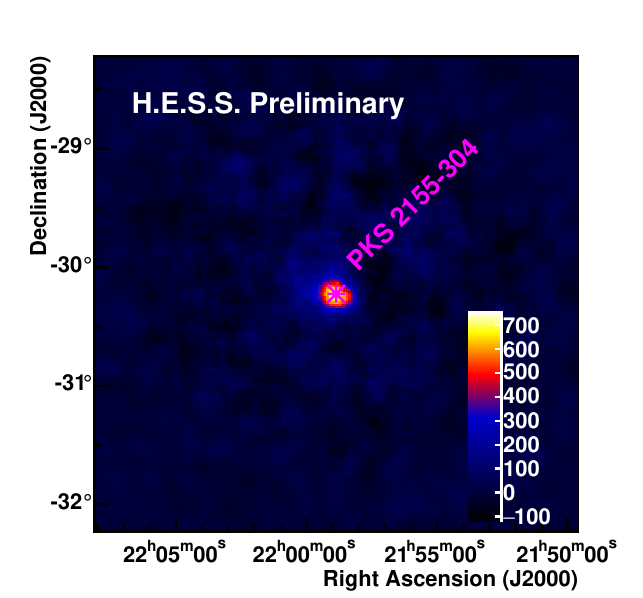}
	\includegraphics[width=0.325\textwidth]{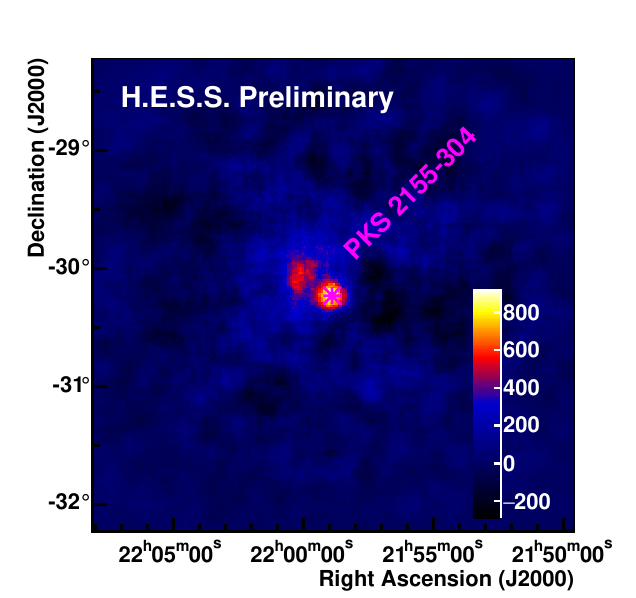}
	\includegraphics[width=0.325\textwidth]{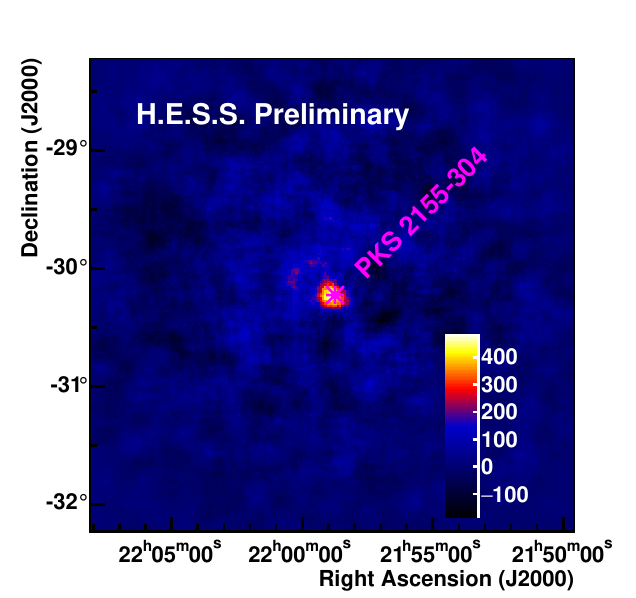}
	\caption{\label{PSK Maps}
		Form left to right, the following cut configurations : $G_{NSB} > 24$, $G_{NSB} > 18$ and $G_{NSB} > 18 + G_{NSB}^{ToM} > 1050$.
		On top, significance distribution of the OFF events (in red) and excess events (in black). On bottom, excess maps of the PKS 2155-304 region.
	}
\end{figure}

\subsection{Application on PKS 2155-304 data}

The detection of the AGN PKS 2155-304 by H.E.S.S. was published in \cite{PKS 2155} based on the 2002 data.
Here the studied data set is constituted of 2014 data passing the standard H.E.S.S. run quality selection. 
It results in a total live time of 8.4 hours of CT5 data. 

The table \ref{PKS table} shows the results of the PKS 2155-304 analysis using the three sets of cuts determined in the precedent sections.  
Lowering the $G_{NSB}$ cut from 24 to 18 increases the number of gamma candidates, but decreases significantly the signal over background ratio in the same time. 
The OFF event significance distribution, presented in the left and middle of the figure \ref{PSK Maps}, enlarges when decrease the $G_{NSB}$ cut value. 
The width of this distribution increases from a perfect 1.0 $\sigma$ to 1.3 $\sigma$.
This behaviour has a direct effect on the excess map of the PKS 2155-304 region shown on left and middle of the figure \ref{PSK Maps}, which presents a not expected bump on the top left of the source, when using the lowest $G_{NSB}$ cut.  

Adding a $G_{NSB}^{ToM}$ cut at 1050 decreases significantly the number of excess events and the source significance, when the signal over background remain more or less stable (table \ref{PKS table}). 
But this cut improves the OFF event significance distribution and the map of the PKS 2155-304 region (right of the figure \ref{PSK Maps}).
The width of the OFF significance distribution decreases from 1.3 $\sigma$ to 1.1 $\sigma$ and the not expected bump on the excess map disappears.

\begin{figure}[tbhp]
	\center
	\includegraphics[width=0.54\textwidth]{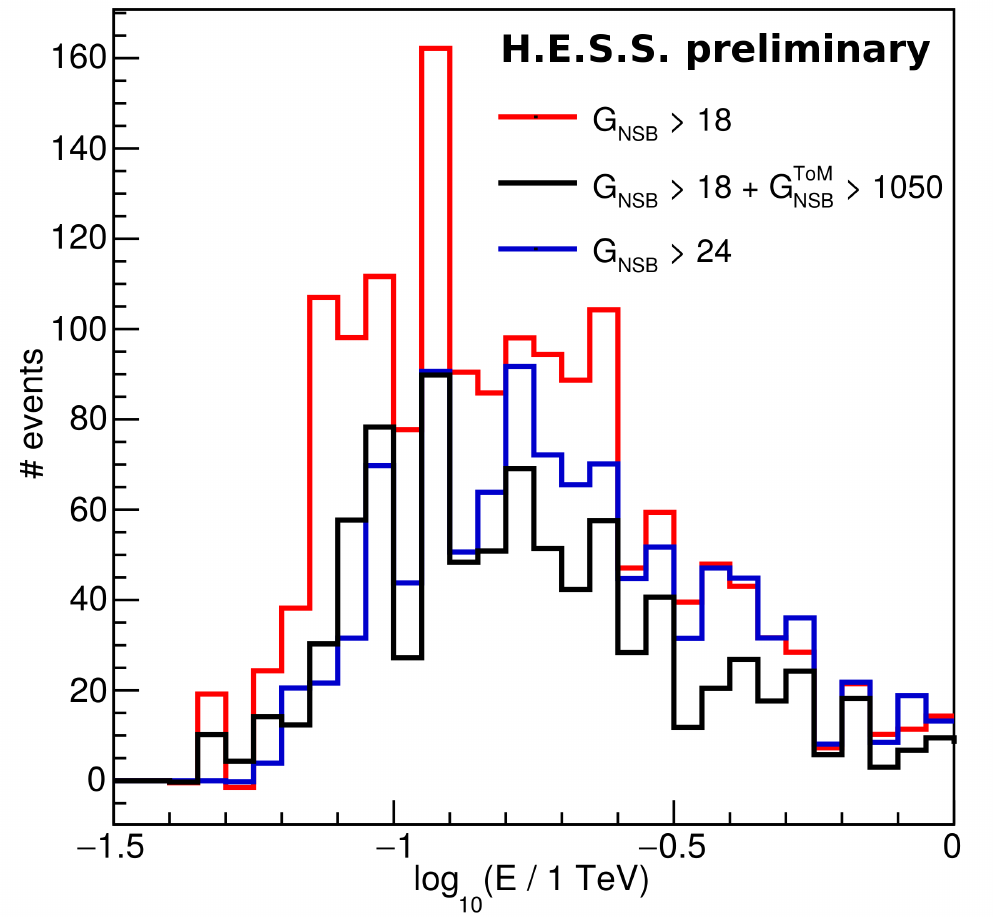}
	\caption{\label{Energy distrib}Distribution of the reconstructed energy of the excess events}
\end{figure}

Finally, the figure \ref{Energy distrib} presents the reconstructed energy distribution of the excess events for the three studied cut configurations.
The decreases of the energy threshold when lowering the $G_{NSB}$ cut from 24 to 18 is once again verified.
Adding the $G_{NSB}^{ToM}$ cut lets the energy distribution shape almost unchanged, so the energy threshold doesn't increase as expected from Monte Carlo simulation (see section \ref{MC section}).
Compared to the $G_{NSB} > 24$ cut (official loosest cut of the Model analysis), the energy distribution using the ToM variable present more excess at low energy and less excess at high energy.
This phenomenon can easily explain the decrease of the significance and number of excess events, since the $G_{NSB}^{ToM}$ cuts more high energy excess events than it gains at low energy.
The time of arrival of the Cherenkov photons emitted by the highest energetic gamma is expected to be wider due to the larger hight of the electromagnetic shower, explaining the loss of the high energetic events when using a $G_{NSB}^{ToM}$ cut.

\section{Conclusion}

A new variable using the Time of Maximum recorded by the fifth H.E.S.S. camera has been introduced.
This variable is constructed to reduced the contamination of shower events with NSB photons.
Tests on Monte Carlo simulations and real data analysis of the PKS 2155-304 regions have shown that it can also be used for hadron rejection and allows to decrease significantly the energy threshold compared to the loosest official H.E.S.S. cut configurations, while keeping control on the background subtraction.
All results presented in this paper are preliminary.


%
%

\end{document}